\documentclass[aip,jap,reprint,nofootinbib]{revtex4-1}
\usepackage{graphicx}% Include figure files
\usepackage{times}
\usepackage{mathptmx}
\usepackage{mathrsfs}
\usepackage{amsmath,amssymb}

\allowdisplaybreaks
\newcommand{\eqhyp}[1]{#1\nobreak\discretionary{}{\hbox{\ensuremath{#1}}}{}}

\begin{document}
\abovedisplayskip=3pt plus 3.0pt minus 3.0pt
\abovedisplayshortskip=0.0pt plus 3.0pt
\belowdisplayskip=3pt plus 3.0pt minus 3.0pt
\belowdisplayshortskip=3pt plus 3pt minus 3pt

\title{\large Superconducting-coil--resistor circuit with electric field quadratic in the current}
\author{\vspace{6pt}N.A. Poklonski}
\email{poklonski@bsu.by}
%\thanks{}
\author{S.Yu. Lopatin}
\affiliation{\vspace{6pt}Belarusian State University, pr. Nezavisimosti 4, 220030 Minsk, Belarus}


\date{\today}

\begin{abstract}
\parbox{138mm}{\vspace{6pt}\raggedright
It is shown for the first time that the observed [Phys. Lett. A \textbf{162}, 105 (1992)] potential difference~$\Phi_\text{t}$ between the resistor and the screen surrounding the circuit is caused by polarization of the resistor because of the kinetic energy of the electrons of the superconducting coil. The~proportionality of $\Phi_\text{t}$ to the square of the current and to the length of the superconducting wire is explained. It is pointed out that measuring $\Phi_\text{t}$ makes it possible to determine the Fermi~quasimomentum of the electrons of a metal resistor.}
\end{abstract}

\keywords{}%Use showkeys class option if keyword display desired

\maketitle %\maketitle must follow title, authors, abstract and \pacs


Edwards \emph{et al.}\cite{1,2} observed the appearance of an electric potential that was quadratic in the current $I$ on a resistor that formed a closed circuit with a superconducting coil relative to a grounded screen (see Fig. 1). They showed experimentally that the potential difference between the center of the resistor $R$ and the screen satisfies $\Phi_\text{t} \propto bI^2$, where $b$ is the length of the superconducting wire. The value of $\Phi_\text{t}$ is virtually independent of the configuration inductance of the coil, $L_\text{c} \approx 800~\mu$H, which is minimized by bifilar winding. When the current is $I = 16$~A and $b \approx 700$~m, for $R = 82~\mu\Omega$, a typical value is $\Phi_\text{t} \approx 5$~mV. The time constant of the circuit is $\tau = L_\text{c}/R \approx 10$~sec. A superconducting wire (NbTi, pure Nb, and also Pb) with a diameter of $2a \approx 127~\mu$m was covered with a copper layer (thickness 19~$\mu$m) whose resistance was much greater than $R$. The coil and the resistor were immersed in liquid helium; the signal wire of the electrometer $V$ was connected to the center of the resistor. The resistor and the electrostatic screen were made from brass.

In principle, a steady-state electric field quadratic in current appears along a rectilinear superconductor of finite length.\cite{3} However, for a closed circuit, it is necessary to take into account the retardation of fields\cite{4} from charges moving with acceleration because of the rotation of their current velocity vector ~this was neglected in Ref.~\onlinecite{5}. When a consistent treatment is used, it is found that there is no electric field around a superconducting circuit with a steady-state current.\cite{6} At the same time, it is shown in Ref.~\onlinecite{7} that, if the ratio of the distance between the ``superconducting'' electrons to the distance between the atomic residues (the ions) is assumed to be equal to $\sqrt{1 - \beta^2}$, where $\beta$ is the ratio of the current velocity of the electrons to the velocity of light, the total electric field is proportional to $\beta^2$ even when it is averaged over a sphere surrounding a circuit with a current. However, this assumption has not been proven.\cite{8,9,10}

The appearance of an electric field proportional to $I^2$ can also be caused by the accelerated motion of charges in the curvilinear section of a conductor,\cite{11} by the redistribution under the action of the intrinsic magnetic field of the electric charge density (the radial Hall effect)\cite{12,13} or of the current in an electrically neutral metal medium (the pinch effect),\cite{14} as well as by the difference of the cross-sectional area of different sections of a conductor (the Bernoulli effect).\cite{13,15} However, quantitative estimates do not allow Edwards \emph{et al.}'s results to be regarded as a manifestation of these effects (see also Refs.~\onlinecite{1} and \onlinecite{2}). This served as the basis for the discussion of Refs.\onlinecite{7,8,9,10} on the relativistic invariance of the equations that describe the total charge of a closed system with a current. (In the recent ``optical'' experiment of Ref.~\onlinecite{16}, the Lorentz transformations were confirmed with an accuracy of $7{\times}10^{-5}$.)

The goal of this paper is to interpret the experiments of Edwards \emph{et al}.\cite{1,2}

We start from the fact that the total energy of a superconducting coil with current $I$ is the sum of the magnetic-field energy $L_\text{c}I^2/2$ and the kinetic energy $K$ of the directed motion of the electrons. For a coil wound with wire $2a$ in diameter and $b$ in length, the kinetic energy of the electrons is\cite{17}
\begin{equation}\label{eq:01}
   K = \frac{L_\text{k}I^2}{2} = \frac{\mu_0\lambda b}{8\pi a}I^2,
\end{equation}
where $L_\text{k}$ is the kinetic inductance, $\lambda \ll a$ is the London current-excitation depth in the wire, and $\mu_0$ is the permeability of free space.

Under the conditions of Edwards \emph{et al}.'s experiment, the total inductance of the coil is $L_\text{t} = L_\text{c} + L_\text{k} \approx L_\text{c}$.

After the resistor is added to the superconducting circuit, the current damps out and an induction electric field appears, aligned with the current. The induction potential difference at the ends of the resistor is $U_\text{i} \approx -L_\text{c}\,\mathrm{d}I/\mathrm{d}t = IR$; the potential of the center of the resistor is $\Phi_\text{i} = U_\text{i}/2$ (here and below, relative to an infinitely remote point, where the potential is set equal to zero). A potential $\Phi_\text{n}$, proportional to the current in the coil, appears at the point where the electrometer is grounded, as a consequence of the mutual inductance of the circuit and the screen. It is clear that $\Phi_\text{n}$ has different magnitudes (and signs) at different sections of the screen because of the vortex character of the currents $I_\text{n}$.

At the initial instant when the resistor is connected into the superconducting circuit, all of its $N$ conduction electrons are displaced by the mean free path $l$ relative to the atomic residues, in the direction opposite the current. As a result, the resistor is polarized, thereby consuming part of the energy stored in the superconducting coil.\cite{18,19} The potential energy of the polarization is $W = EP/2$, where $E$ is the electric field, $P = elN$ is the dipole moment, and $e > 0$ is the elementary charge. The polarization potential difference on the ends of a resistor of length $d$ is $U_\text{p} = Ed$; the potential of the center of the resistor is $\Phi_\text{p} = U_\text{p}/2$.

\begin{figure}%[!h]
\noindent\hfil\includegraphics[width=.9\columnwidth]{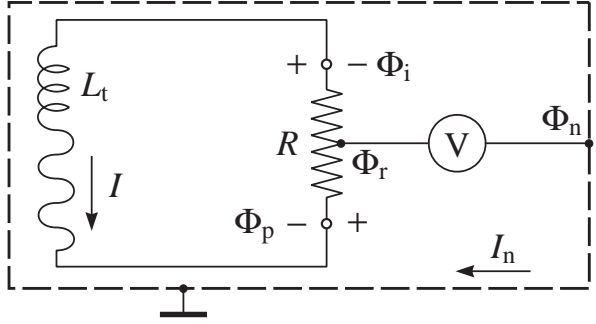}
\caption{Schematic representation of the measurements of Edwards \emph{et al}.\cite{1,2}}\label{fig:01}
\end{figure}

Since the energy $W$ consumed by the polarization of the resistor is limited to $K$, from the relationship $U_\text{p}P/2d = L_\text{k}I^2/2$ and using Eq.~\eqref{eq:01}, we get the estimate
\begin{equation}\label{eq:02}
   \Phi_\text{p} = \frac{L_\text{k}I^2d}{2elN} = \frac{\mu_0\lambda b}{8\pi ealSn}I^2,
\end{equation}
where $n = N/Sd$ is the density of conduction electrons in a resistor with cross-sectional area $S$ and length $d$.

It follows from Eq.~\eqref{eq:02} that the polarization potential $\Phi_\text{p}$ is proportional to the square of the current and the length of the superconducting section of the circuit but is independent of the length of the resistor.

At liquid-helium temperature, the contribution of the additional resistance of the resistor--superconductor junction (the critical temperature is $\approx$\,10~K) to creating a potential difference at the ends of the resistor is negligible.\cite{17}

Provided that $e(U_\text{p} - IR) = 2e(\Phi_\text{p} - \Phi_\text{i})$ remains less than the energy gap of the superconductor ($\approx$\,3~meV), the current in the circuit ceases.

The potential difference between the resistor and the screen measured in Edwards \emph{et al}.'s experiments is thus $\Phi_\text{t} \eqhyp= \Phi_\text{r} \eqhyp- \Phi_\text{n}$, where $\Phi_\text{r} = \Phi_\text{p} - \Phi_\text{i}$ is the potential of the center of the resistor (the point where the signal wire is connected), and $\Phi_\text{n}$ is the potential of the grounding point of the electrometer.

From the parameters of a brass resistor, $R = 82~\mu\Omega$, $d = 14$~mm, and $S \approx 0.1$~cm$^2$, we find its conductivity $\sigma \approx 1.6{\times}10^5~\Omega^{-1}{\cdot}$cm$^{-1}$. According to the data of Ref.~\onlinecite{20}, the density of conduction electrons in brass (Cu$_{1-x}$Zn$_x$) with an atomic density of 8.65~g$\cdot$cm$^{-3}$ for $x \approx 0.2$ is $n \eqhyp\approx 9.8{\times}10^{22}$~cm$^{-3}$. At liquid-helium temperature, the mean free path $l = p_\text{F}\sigma/e^2n$ of the conduction electrons is determined by the quasi-momentum $p_\text{F} = \hbar(3\pi^2n)^{1/3}$ at the Fermi surface. Substituting the values $\lambda \approx 38$~nm (for Nb and Pb) and $l \approx 10$~nm (for brass) into Eq.~\eqref{eq:02}, with $I = 16$~A gives $\Phi_\text{p} \approx 4$~mV and $\Phi_\text{i} \approx 0.65$~mV. The potential difference $\Phi_\text{t} \approx \Phi_\text{r} = \Phi_\text{p} - \Phi_\text{i} \approx 3.3$~mV (for $\Phi_\text{n} = 0$) is comparable with the experimental value.

The sign of $\Phi_\text{t} = \Phi_\text{r} - \Phi_\text{n}$ must be independent of the direction of the current in the circuit for a given electrical configuration.

Note that, when $l \propto p_\text{F}/n$, $\Phi_\text{p} \propto \lambda/p_\text{F}$ follows from Eq.~\eqref{eq:02}; i.e., the Fermi quasi-momentum $p_\text{F}$ of the electrons in the resistor can be determined by measuring $\Phi_\text{t} \approx \Phi_\text{p}$.

Thus, the superconducting-coil--resistor circuit in Edwards \emph{et al}.'s experiments\cite{1,2} has an electric field that is quadratic in current because the polarization electric field of the resistor predominates over the induction field. In a wellknown sense, Edwards \emph{et al}.'s experiments supplement the experiments of Tolman \emph{et al}. (see Ref.~\onlinecite{21}), who measured a damped current in a closed circuit, caused by the motion of the conduction electrons due to inertia after the removal of a rotating nonsuperconducting coil.

We are grateful to I. Z. Rutkovski\u\i, G. S. Kembrovski\u\i, and V. V. Mityanok for discussions.

This work was carried out within the Low-Dimension Systems Program of the Ministry of Education of the Republic of Belarus.

%\newpage
%~\\
%\newpage
%\bibliography{poklonski}

%
\end{document}